\documentclass[aps,reprint,pre,showkeys,superscriptaddress]{revtex4-1}

\pdfoutput=1
\usepackage{graphicx}
\usepackage{amsmath}
\usepackage{natbib}
\usepackage{epstopdf}

\begin{document}

\title{Scaffold-mediated Nucleation of Protein Signaling Complexes: Elementary Principles}

\author{Jin Yang}
\affiliation{Chinese Academy of Sciences -- Max Planck Society Partner Institute for Computational Biology, Shanghai Institutes for Biological Sciences, 320 Yue Yang Road, Shanghai 200031, China. E-Mail: jinyang2004@gmail.com}

\author{William S. Hlavacek}
\affiliation{Theoretical Biology and Biophysics Group, Theoretical Division and Center for Nonlinear Studies, Los Alamos National Laboratory, Los Alamos, NM 87545, USA}
\affiliation{Department of Biology, University of New Mexico, Albuquerque, NM 87131, USA. E-Mail: wish@lanl.gov}

\begin{abstract}
Proteins with multiple binding sites play important roles in cell signaling systems
by nucleating protein complexes in which, for example, enzymes and substrates are
co-localized. Proteins that specialize in this function are called by a variety
names, including adapter, linker and scaffold. Scaffold-mediated nucleation of
protein complexes can be either constitutive or induced. Induced nucleation is
commonly mediated by a docking site on a scaffold that is activated by
phosphorylation.  Here, by considering minimalist mathematical
models, which recapitulate scaffold effects seen in more mechanistically
detailed models, we obtain analytical and numerical results that provide
insights into scaffold function.  These results elucidate how recruitment of a
pair of ligands to a scaffold depends on the concentrations of the ligands, on
the binding constants for ligand-scaffold interactions, on binding
cooperativity, and on the milieu of the scaffold, as ligand recruitment is
affected by competitive ligands and decoy receptors.  For the case of a bivalent
scaffold, we obtain an expression for the unique scaffold concentration that
maximally recruits a pair of monovalent ligands.  Through simulations, we
demonstrate that a bivalent scaffold can nucleate distinct sets of ligands to
equivalent extents when the scaffold is present at different concentrations. 
Thus, the function of a scaffold can potentially change qualitatively with a
change in copy number.  We also demonstrate how a scaffold can change the
catalytic efficiency of an enzyme and the sensitivity of the rate of reaction to
substrate concentration. The results presented here should be useful for
understanding scaffold function and for engineering scaffolds to have desired
properties. 
\end{abstract}


\maketitle

\section{Introduction}
Proteins involved in cell signaling typically possess multiple binding
sites, and as a result, a common feature of these proteins is their ability to
interact with several binding partners to form heterogeneous protein
complexes~\cite{hlavacek2003ccs}. Some proteins specialize in
the nucleation of protein complexes~\cite{paw:97}.  For example, Ste5 is a
well-studied scaffold protein that acts in a yeast MAP kinase cascade; it binds all three kinases of the cascade, Ste11,
Ste7, and Fus3~\cite{Elion01}.  The interaction of Fus3 with Ste5 is modulated by multisite phosphorylation of Ste5 \cite{malleshaiah2010}.  One function of Ste5 is to co-localize
enzyme-substrate pairs (e.g., Ste11 and Ste7) in the cascade~\cite{Park:03},
which contributes to switch-like cellular responses to signals
\cite{malleshaiah2010}.   Two well-known examples of scaffold-like proteins
that nucleate protein complexes in mammalian signaling systems are the Grb2 adapter
protein in the epidermal growth factor receptor (EGFR)
pathway~\cite{Rozakis-Adcock93} and the LAT linker protein in the T cell
receptor (TCR) pathway~\cite{Zhang98, samelson2002stm,bwr:02}.  The binding sites in LAT are activated by tyrosine phosphorylation, and as a result, the valence of LAT and the abundance of functional LAT binding sites can be quickly changed by changes in kinase and phosphatase activities, with these changes serving to modulate the effect of LAT on signaling \cite{nag2009biophysj}.  In addition to
proteins with dedicated scaffold functions, proteins that have catalytic
activities and protein complexes, such as ligand-induced receptor dimers,  can
also function as scaffolds, i.e., take on the function of facilitating the
formation of multicomponent complexes through multivalent binding.  

Many functions have been ascribed to scaffold
proteins~\cite{paw:97,bwr:02,brw:00,pawson2007,zeke2009,good2011}. For example, scaffolds have been suggested to control
the specificities of enzymes, coordinate spatiotemporal aspects of signaling,
and amplify or attenuate signals. To better understand the
mechanisms by which scaffolds influence cell signaling, researchers have formulated mathematical models
to study the effects of scaffolds on specific signaling
systems and to study the generic properties of
scaffolds~\cite{lay:97,lev:00,hein:02,sig:02,prehoda2002spi,locasale2007spc,
locasale2008rsd,malleshaiah2010,thalhauser2010}. The ability of a scaffold to either amplify or attenuate
signaling was demonstrated  by Levchenko et al.~\cite{lev:00} using a
mathematical  model for a MAP kinase cascade. In this model, scaffold
enhancement of signaling results from scaffold-mediated nucleation of an enzyme
with its substrate, whereas scaffold inhibition of signaling results from excess
scaffold, which separates enzyme and substrate into distinct enzyme-scaffold and substrate-scaffold complexes.  This effect is equivalent to the
well-known prozone effect in antibody-antigen reactions~\cite{lay:97}.  Heinrich
et al.~\cite{hein:02} further studied these effects of scaffolds on protein
kinase cascades. More recently, Locasale et al.~\cite{locasale2007spc} studied how scaffolds
affect competition between phosphatases and kinases, finding  that scaffolds can
have positive and negative effects on signaling for reasons related to the
kinetics of signaling events.  

Modeling has improved our understanding of scaffold function, but many of the reported modeling studies of scaffold function have been based solely on simulations, which makes the results dependent on the parameter values considered in simulations.  In addition, many of the models used to study scaffolds incorporate mechanistic details that are likely to be extraneous for understanding scaffold function broadly.  Greater theoretical understanding of the design principles of scaffolds could perhaps be obtained by focusing on minimalist models that are relatively easy to analyze. With a better understanding of scaffold design principles, we can hope to manipulate scaffold function through precise tuning of scaffold and system properties.  Manipulating the properties of a scaffold or its milieu to alter the behavior of a cell signaling system is known to be feasible \cite{Park:03,bashor2008ues,chapman2009}. For example, Lim and co-workers have
demonstrated that the response of a cell to a signal can be changed
qualitatively by altering the binding specificities of a
scaffold~\cite{Park:03}.  

Here, by considering minimalist models for
multivalent scaffold-ligand interaction, we investigate the recruitment of ligands to a
scaffold and the effects of a scaffold on an enzymatic reaction. We obtain a number of analytical and numerical results, mostly for a model that characterizes the equilibrium formation of a ternary complex composed of a scaffold and two binding partners.  In this ternary-complex model, the scaffold is bivalent and and its monovalent binding partners are taken to be an enzyme-substrate pair.  Each
binding partner, or ligand, interacts with one of the two binding sites of the
scaffold.  We
evaluate the relevance of the ternary-complex model through comparisons with
more mechanistically detailed models for cell signaling systems that involve scaffolds.  The ternary-complex model is similar in mathematical form to a variety of models that have been used to study receptor signaling \cite{jacobs1976,delean1980} and multivalent ligand-receptor binding \cite{perelson1980,sulzer1996,mack2008exact}.

\begin{figure}
\centering
\includegraphics[scale=1.0]{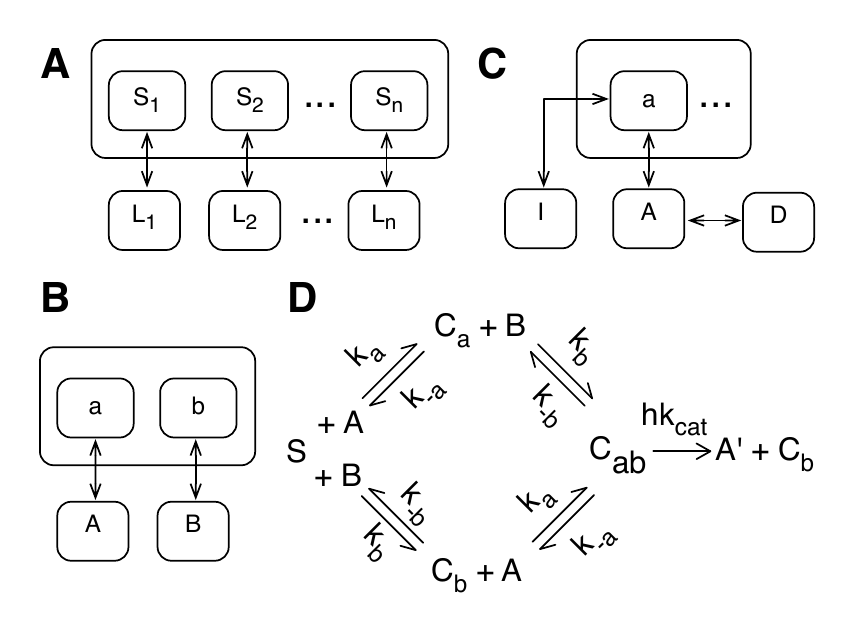}
\caption{\label{fig:1} Minimalist models for ligand-scaffold interactions. (A) A
scaffold with $n$ sites ($S_1,\ldots,S_n$).  Each site $S_i$ interacts with a distinct ligand $L_i$.  (B) Ternary-complex model. A bivalent scaffold with sites $a$ and $b$ interacts with
monovalent ligands $A$ and $B$, respectively.  (C) A scaffold site $a$ interacts with a ligand $A$ and a competitive inhibitor $I$.  The ligand $A$ interacts with a decoy receptor $D$.  (D) A reaction scheme for a scaffold-mediated enzymatic reaction that produces a product $A^{\prime}$.  The substrate is scaffold ligand $A$ and the enzyme is scaffold ligand $B$.  See text for discussion.}
\end{figure}

The remainder of this report is organized as follows.  In Section~\ref{sec:theor}, we introduce models and present analytical results relevant for understanding the effects of scaffold valence, negative and positive cooperativity in binding of multiple ligands to a scaffold, competitive inhibitors of ligand-scaffold interactions, and decoy receptors.  We also develop an approximate rate law for scaffold-mediated enzyme kinetics.  Our most significant analytical result is a design equation that characterizes how the scaffold concentration that maximally nucleates a ternary complex depends on equilibrium binding constants and ligand concentrations.  Interestingly, this scaffold concentration is independent of cooperativity in ligand binding.  In Section~\ref{sec:res}, we present numerical results that complement the analytical results.  The numerical results include a demonstration that our simple ternary-complex model recapitulates scaffold effects seen in more mechanistically detailed models and a demonstration that a single scaffold can nucleate distinct signaling complexes at different scaffold concentrations.  We conclude with a brief discussion (Section~\ref{sec:discuss}).

\section{Models and Analytical Results}\label{sec:theor}

The models that we will consider are illustrated in Fig.~\ref{fig:1} and described below, along with various analytical results relevant for understanding scaffold function.

\subsection{Multivalent scaffold with independent binding sites}

We consider a scaffold protein with $n \geq 2$ binding sites.  Each scaffold binding site $i$ interacts independently with a distinct monovalent ligand $i$, as illustrated in Fig.~\ref{fig:1}(A).   Thus, we consider the following reactions: 
\begin{equation}\label{eq:1}
S_i + L_i \rightleftharpoons B_i, \ \ i=1,\ldots, n \ , 
\end{equation}
where $S_i$ represents free site $i$, $L_i$ represents free ligand $i$, and $B_i$ represents bound site $i$ (or equivalently, bound ligand $i$).  Using the same symbols for chemical species in Eq.~(\ref{eq:1}) for the corresponding concentrations, we can write the following equilibrium relations, which are derived from the law of mass action:
\begin{equation}\label{eq:binding}
K_iB_i = S_iL_i,  \ \ i=1,\ldots,n \ ,
\end{equation}
where $K_i$ is an equilibrium dissociation constant.  Assuming conservation of mass, such that $S_0=S_i+B_i$ and $L_{i_0}=L_i+B_i$, where $S_0$ is the total concentration of scaffold protein and $L_{i_0}$ is the total concentration of ligand $i$, we find from Eq.~(\ref{eq:binding}) that
\begin{equation}\label{eq:quadratic}
B_i^2-(S_0+L_{i_0}+K_i)B_i+S_0L_{i_0}=0 \ . 
\end{equation}
From the quadratic formula and the physical constraint that $B_i < \min(L_{i_0},S_0)$, it further follows that
\begin{equation}\label{eq:bc}
B_i =\frac{S_0+L_{i_0}+K_{i}-\sqrt{(S_0+L_{i_0}+K_{i})^2-4S_0L_{i_0}}}{2} \ .
\end{equation}
This equation and the mass conservation relations given above can be used to completely determine the equilibrium state of the system described in Fig.~\ref{fig:1}(A).

We will now show that there is a unique scaffold concentration $S_0^{\rm opt}$ that maximizes $C_n$, the equilibrium concentration of scaffold protein with all $n$ sites occupied.   The analytical results that follow complement numerical results obtained in earlier modeling studies of scaffold function~\cite{brw:00,lay:97,lev:00,hein:02}.  

We note that the equilibrium quantity $B_i/S_0$ can be interpreted as the probability that scaffold binding site $i$ is occupied at equilibrium.  Likewise, $C_n/S_0$ can be interpreted as the joint probability that all $n$ scaffold binding sites are occupied. Thus, because each scaffold binding site is assumed to interact with its cognate ligand independently, we can write
\begin{equation}\label{eq:complex}
C_n=S_0^{1-n} \prod_{i=1}^nB_i \ .
\end{equation}
Trivially, for the case $S_0=0$, $C_n=0$.  In other words, there is no complex in the absence of scaffold.  Also, from Eq.~(\ref{eq:complex}), we can see that $\lim_{S_0\to\infty}C_n =0$, because each $B_i$ is finite.  In fact $B_i$ can be no greater than $L_{i_0}$.  From inspection of Eqs.~(\ref{eq:bc}) and (\ref{eq:complex}), one can see that $C_n$ is a continuous and differential function of $S_0$.  Thus, from the mean value theorem, the derivative of $C_n$ with respect to $S_0$ must vanish at some finite value of $S_0$, which we will denote as $S_0^{\rm opt}$.  In other words,
\begin{equation}\label{eq:nesscon}
\left . \frac{dC_n}{dS_0}\right |_{S_0=S_0^{\rm opt}}=0 \ .
\end{equation}
Because $S_0^{\rm opt}$ is unique, as we will show below, and $C_n$ is positive for all finite $S_0$ (on physical grounds), $C_n$ is maximum at $S_0=S_0^{\rm opt}$.

To see that $S_0^{\rm opt}$ is unique, consider the following expression, which is derived from Eq.~(\ref{eq:complex}) by differentiating $C_n$ with respect to $S_0$:
\begin{equation}\label{eq:dcds}
g_n \equiv \frac{S_0}{C_n}\frac{dC_n}{dS_0} = \sum_{i=1}^n b_i - (n-1) \ ,
\end{equation}
where
\begin{equation}\label{eq:bi}
b_i \equiv \frac{S_0}{B_i} \frac{dB_i}{dS_0} = \frac{S_0(B_i-L_{i_0})}{B_i^2-S_0L_{i_0}} \ .
\end{equation}
Note that $g_n$ is the slope of the plot of $\log C_n$ vs. $\log S_0$.  The expression for $b_i$ was obtained by differentiating each term in Eq.~(\ref{eq:quadratic}) with respect to $S_0$ and then solving for $dB_i/dS_0$.  From inspection of Eq.~(\ref{eq:bi}), one can see that $0 < b_i < 1$ for finite $S_0$.   In fact, it can be shown that $\lim_{S_0 \to 0} b_i = 1$ and $\lim_{S_0 \to \infty} b_i = 0$.  From these considerations and inspection of Eq.~(\ref{eq:dcds}), one can see that $S_0^{\rm opt}$ is unique if $b_i$  decreases monotonically as $S_0$ increases for all $i$.  Differentiating the expression for $b_i$ given in Eq.~(\ref{eq:bi}) with respect to $S_0$, we find
\begin{equation}\label{eq:concave}
\frac{db_i}{dS_0} =\frac{B_i(B_i-L_{i_0})(B_i-S_0)(B^2_i+S_0L_{i_0})}{(B_i^2-S_0L_{i_0}
)^3} < 0 \ .
\end{equation}
Note that the terms $B_i-L_{i_0}$, $B_i - S_0$, and $B_i^2 - S_0 L_{i_0}$ are all negative. Thus, each $b_i$ is a monotonically decreasing function, as is $g_n$ by extension, and as a result, $S_0^{\rm opt}$ is unique.   

The results presented above provide insight into the shape of the log-log scaffold dose-response curve, meaning the plot of $\log C_n$ vs. $\log S_0$.  Recall that the slope of this curve, which is concave, is given by Eq.~(\ref{eq:dcds}).  As can be seen from Eq.~(\ref{eq:dcds}), for $S_0<S_0^{\rm opt}$, $1> g_n > 0$, whereas for $S_0>S_0^{\rm opt}$, $0 > g_n > 1- n$.  For $n=2$, $|g_2|$ approaches 1 as $S_0$ approaches either 0 or $\infty$.  However, for $n \geq 3$, the two asymptotic limits of $|g_n|$ are different, with the curve becoming steeper at some point in the excess scaffold regime ($S_0 > S_0^{\rm opt}$) than at any point in the regime where $S_0 < S_0^{\rm opt}$.  Thus, the log-log scaffold dose-response curve is asymmetric.

To gain insight into the effect of scaffold valence on scaffold function, measured by nucleation of a complex in which all scaffold binding sites are occupied, let us consider two scaffolds in the same milieu with the same properties except that one scaffold has an extra binding site $n$ and cognate binding partner $L_n$.  We impose this equivalence constraint to ensure a fair comparison \cite{savageau1972,savageau1976}.  We will denote the valences of the two scaffolds as $m \geq 2$ and $n=m+1$.  From Eq.~(\ref{eq:complex}) and our equivalence constraint, it follows that $C_n = C_m B_n/S_0$ and further that $C_n<C_m$ for any given scaffold concentration.  Thus, with all factors being as equal as possible, scaffold-mediated nucleation of $C_n$ is less efficient than nucleation of $C_m$. Another inherent effect of valence on scaffold function is greater sensitivity of $C_n$ to changes in scaffold concentration in the excess scaffold regime. From Eq.~(\ref{eq:dcds}), for $S_0$ sufficiently large such that $g_m<0$ and $g_n<0$, it can be seen that $g_n$ is larger in magnitude than $g_m$.  This result is consistent with our earlier discussion of asymmetry in the log-log scaffold dose-response curve.  

\subsection{Bivalent scaffold with cooperative ligand binding}

We will now consider a bivalent scaffold that interacts with a pair of monovalent ligands $A$ and $B$ (Fig.~\ref{fig:1}(B)).  Ligand $A$ interacts with a scaffold binding site $a$, and ligand $B$ interacts with a scaffold binding site $b$.  Equation~(\ref{eq:1}) with $n=2$ applies, but we will not use the nomenclature of Eq.~(\ref{eq:1}) in our treatment of this special case, which because of its simplicity will allow us to obtain an analytical expression for the scaffold concentration that maximizes formation of the ternary complex composed of $A$, $B$ and the scaffold.  Nucleation of this complex can be taken as a measure of the functional activity of the scaffold if $A$ and $B$ constitute an enzyme and substrate or if instead the two ligands constitute functionally related enzymes recruited to the scaffold to act in concert, as in substrate channeling \cite{dueber2009}.  We will relax the assumption that ligands interact with the scaffold independently and allow for cooperative binding.  Thus, using the law of mass action, we write the following equilibrium relations:
\begin{eqnarray}
K_aC_a &=& S_f A_f \label{eq:sa} \ ,\\
K_bC_b &=& S_f B_f \label{eq:sb}\ ,\\
K_aC_{ab} &=& \phi A_f C_b \label{eq:cab} \ .
\end{eqnarray}
where $S_f$ is the concentration of free scaffold, $A_f$ and $B_f$ are the concentrations of free ligands $A$ and $B$, $K_a$ is an equilibrium dissociation constant that characterizes $a$-$A$ interaction, $C_a$ is the concentration of the binary complex composed of $A$ and the scaffold, $K_b$ is an equilibrium dissociation constant that characterizes $b$-$B$ interaction, $C_b$ is the concentration of the binary complex composed of $B$ and the scaffold, and $C_{ab}$ is the concentration of the ternary complex composed of $A$, $B$, and the scaffold. The factor $\phi$, which is positive and dimensionless, is introduced to characterize cooperativity in binding of $A$ and $B$ to the scaffold: $\phi<1$ indicates negative cooperativity, $\phi>1$ indicates positive cooperativity, and $\phi=1$ indicates that the ligands bind the scaffold independently.  We will assume conservation of mass, such that the following relations hold:
\begin{eqnarray}
S_0&=&S_f+C_a+C_b+C_{ab} \label{eq:s0}\ , \\
A_0&=& A_f +C_a+C_{ab} \label{eq:a0} \ ,\\
B_0&=& B_f +C_b+C_{ab} \label{eq:b0}\ .
\end{eqnarray}
where $S_0$ is the total scaffold concentration, $A_0$ is the total concentration of ligand $A$, and $B_0$ is the total concentration of ligand $B$.  

We will now derive an expression for $S_0^{\rm opt}$, the total scaffold concentration that maximizes $C_{ab}$, or equivalently, the total scaffold concentration for which $dC_{ab}/dS_0=0$.  From Eqs.~(\ref{eq:sa}) and~(\ref{eq:cab}), we find
\begin{equation}\label{eq:cab2}
 C_{ab}= \phi \frac{C_a C_b}{S_f} \ . 
\end{equation}
We note that, for $\phi=1$, $C_{ab}=C_aC_b/S_f = (C_a + C_{ab})(C_b + C_{ab})/S_0$, which is equivalent to Eq.~(\ref{eq:complex}) with $n=2$. From Eqs.~(\ref{eq:sa}) and~(\ref{eq:a0}), we find
\begin{equation}\label{eq:sa2}
C_a=\frac{S_f(A_0-C_{ab})}{S_f+K_a} \ .
\end{equation}
Similarly, from Eqs.~(\ref{eq:sb}) and (\ref{eq:b0}), we find
\begin{equation}\label{eq:sb2}
C_b=\frac{S_f(B_0-C_{ab})}{S_f+K_b} \ .
\end{equation}
From Eqs.~(\ref{eq:cab2})--(\ref{eq:sb2}), we find 
\begin{equation}\label{eq:cab3}
 C_{ab}^2- R C_{ab}+A_0B_0=0 \ ,
\end{equation}
where
\begin{equation}\label{eq:reqn}
R \equiv A_0+B_0+(S_f+K_a)(S_f+K_b)/(\phi S_f) \ .
\end{equation}
Differentiating each term in Eq.~(\ref{eq:cab3}) with respect to $S_0$, we find
\begin{equation}
\frac{dC_{ab}}{dS_0} = \left( \frac{C_{ab}\phi^{-1}}{2C_{ab}-R} \right) \left( 1 - \frac{K_aK_b}{S_f^2} \right) \frac{dS_f}{dS_0} \ .
\end{equation}
From this equation, one can see that the derivative $dC_{ab}/dS_0$ uniquely vanishes at $S_f = \sqrt{K_aK_b}$.  Thus, $S_0^{\rm opt}$ is the value of $S_0$ at which $S_f=\sqrt{K_aK_b}$.  Using Eqs.~(\ref{eq:sa2}) and (\ref{eq:sb2}), we can rewrite Eq.~(\ref{eq:s0}) as follows:
\begin{equation} \label{eq:conss2}
S_0 = S_f + \frac{S_fA_0}{S_f+K_a} + \frac{S_fB_0}{S_f+K_b} + \left( 1 - \frac{S_f}{S_f+K_a} - \frac{S_f}{S_f+K_b} \right) C_{ab} \ .
\end{equation}
Substituting $\sqrt{K_aK_b}$ for $S_f$ in Eq.~(\ref{eq:conss2}), we find
\begin{equation}\label{eq:optim}
S_0^{\rm opt} = \sqrt{K_aK_b} + \frac{A_0}{1 + \sqrt{K_a/K_b}} + \frac{B_0}{1 + \sqrt{K_b/K_a}} \ .
\end{equation}
Notably, $S_0^{\rm opt}$ is independent of the cooperativity factor $\phi$.  For the special case $K_a=K_b=K_D$, $S_0^{\rm opt}= K_D + (A_0 + B_0)/2$.  For the special case $A_0=B_0=L_0$, $S_0^{\rm opt}= \sqrt{K_aK_b} + L_0$.

It is straightforward to find an expression for the maximum concentration of scaffold-nucleated ternary complex, which we will denote as $C_{ab}^{\rm max}$.  We simply use the quadratic formula to solve Eq.~(\ref{eq:cab3}) for the case where $R=A_0 + B_0 + \phi^{-1} (\sqrt{K_a} + \sqrt{K_b})^2$.  This expression for $R$ is obtained by substituting $\sqrt{K_aK_b}$ for $S_f$ in Eq.~(\ref{eq:reqn}).  Unlike $S_0^{\rm opt}$, $C_{ab}^{\rm max}$ depends on $\phi$ (because $R$ depends on $\phi$).  For the special case where $\phi=1$, $K_a=K_b=K_D$ and $A_0=B_0=L_0$, we find $C_{ab}^{\rm max} = (\sqrt{K_D+L_0}-\sqrt{K_D})^2$.  

\subsection{Effect of a competitor or decoy receptor}

The \textit{in vivo} milieu of a scaffold is complex.  A given binding site on the scaffold may interact with a specific ligand as well as a number of competitors.  Likewise, the ligand may interact a number of binding partners besides the scaffold. To study the effects of the milieu on the function of a scaffold, we consider the simple scenario illustrated in Fig.~\ref{fig:1}(C). A ligand $A$ interacts with a scaffold binding site $a$ in competition with an inhibitor $I$.  The ligand also interacts with a decoy receptor $D$.  Thus, we consider the following equilibrium relations:
\begin{eqnarray}
K_a B_a & = & S_a A_f \label{eq:cd1} \ , \\ 
K_c B_c & = & S_a I_f  \label{eq:cd2} \ , \\
K_d B_d & = & D_f A_f \label{eq:cd3} \ ,
\end{eqnarray}
where $S_a$ is the concentration of free scaffold binding site $a$, $A_f$ is the concentration of free ligand $A$, $B_a$ is the concentration of scaffold bound to ligand $A$, $I_f$ is the concentration of free inhibitor $I$, $B_c$ is the concentration of scaffold bound to competitive inhibitor $I$, $D_f$ is the concentration of free decoy receptor $D$, and $B_d$ is the concentration of ligand $A$ bound to decoy receptor $D$. The equilibrium dissociation constants $K_a$, $K_c$ and $K_d$ characterize $a$-$A$ interaction, $a$-$I$ interaction and $A$-$D$ interaction, respectively.  We will assume that the following mass conservation relations hold: $S_0 = S_a + B_a + B_c$, where $S_0$ is the total scaffold concentration, and $A_0 = A_f + B_a + B_d$, where $A_0$ is the total ligand concentration.

To quantify the effect of a competitive inhibitor or decoy receptor on interaction between a scaffold binding site and a ligand, we can find the apparent equilibrium dissociation constant $K_a^{\prime}$ satisfying the relation $K_a^{\prime} B_a = S_a^{\prime} A_f^{\prime}$ (cf. Eq.~(\ref{eq:cd1})), where $S_a^{\prime} \equiv S_0 - B_a = S_a + B_c$ is the apparent concentration of free scaffold binding site $a$ and $A_f^{\prime} \equiv A_0 - B_a = A_f + B_d$ is the apparent concentration of free ligand $A$.  From these considerations and Eqs.~(\ref{eq:cd1})--(\ref{eq:cd3}), we find
\begin{equation}\label{eq:app}
K_a^{\prime}= K_a\left(1+\frac{I_f}{K_c}\right)\left(1+\frac{D_f}{K_d}\right) \ .
\end{equation}
Thus, a competitive inhibitor or decoy receptor simply has the effect of reducing the apparent affinity of a ligand for its scaffold binding site.  

In the previous section, we provided equations that elucidate how $S_0^{\rm opt}$ and $C_{ab}^{\rm max}$ depend on binding constants.  Given Eq.~(\ref{eq:app}), we can now interpret these equations as also defining how competitive inhibitors and decoy receptors in the milieu of a scaffold affect $S_0^{\rm opt}$ and $C_{ab}^{\rm max}$.  A competitive inhibitor or decoy receptor can shift the total scaffold concentration that maximally nucleates a signaling complex.  For example, under certain conditions, a competitive inhibitor can promote scaffold-mediated nucleation of a signaling complex by reversing the effect of excess scaffold (i.e., the effect of $S_0>S_0^{\rm opt}$). Thus, the function of a scaffold might be controlled by regulating the expression levels of competitive inhibitors and decoy receptors.  It should also be noted that competitive inhibitors and decoy receptors can obscure the relationship between $S_0^{\rm opt}$ \textit{in vivo} and binding parameters determined \textit{in vitro}.

\subsection{Effect of a scaffold on the rate of an enzyme-catalyzed reaction} 

To study the effect of a scaffold on the rate of an enzyme-catalyzed reaction, we consider the scheme presented in Fig.~\ref{fig:1}(D). This scheme represents a simple extension of the classical Michaelis-Menten reaction scheme:
\begin{equation}
A+ B \underset{k_{-1}}{\overset{k_1}\rightleftharpoons} 
(AB) \overset{k_{\rm cat}}\rightarrow A^{\prime} + B \  ,
\end{equation}
where $A$ is a substrate, $B$ is an enzyme, $(AB)$ is an enzyme-substrate complex, $A^{\prime}$ is a product, and $k_1$, $k_{-1}$ and $k_{\rm cat}$ are rate constants. As is well known, a pseudo steady-state assumption leads to the following approximate expression for the rate of product formation $V \equiv k_{\rm cat} (AB)$:
\begin{equation}\label{eq:MM}
V = \frac{k_{\rm cat} B_0 A_f}{K_m+A_f} \ , 
\end{equation}
where $A_f$ is the concentration of free substrate, $B_0$ is the total concentration of enzyme, and $K_m=(k_{-1}+k_{\rm cat})/{k_1}$.  For simplicity, we will assume that $k_{\rm cat} \ll k_{-1}$ such that $K_m \approx k_{-1}/k_1$.  When $K_m$ is set equal to $k_{-1}/k_1$, it can be said that Eq.~(\ref{eq:MM}) depends on a rapid equilibrium assumption. 

To obtain an expression analogous to that of Eq.~(\ref{eq:MM}) for the reaction scheme of Fig.~\ref{fig:1}(D), we will assume that $V^{\prime}$, the rate of product formation when an enzyme and substrate are co-localized on a scaffold, is proportional the concentration of the ternary complex of the scaffold, enzyme and substrate, which we will denote as $C_{ab}$.  In other words, we will assume that $V^{\prime} \equiv h k_{\rm cat} C_{ab}= h k_{\rm cat} (B_0-B_f-C_b)$, where $h \geq 0$ is a constant introduced to characterize the effect on reactivity of scaffold-controlled positioning of the enzyme and substrate.  Positioning/orientation of an enzyme-substrate pair is a known function of some scaffolds \cite{good2011}.  Note that we have assumed conservation of enzyme, such that $B_0 = B_f + C_b + C_{ab}$, where $B_0$ is the total concentration of enzyme, $B_f$ is the concentration of free enzyme, and $C_b$ is the concentration of the binary complex of enzyme and scaffold (Fig.~\ref{fig:1}(D)).  Under a rapid equilibrium assumption, we take the following relations to hold: $K_bC_b = S_fB_f$ and $K_aC_{ab}=A_f C_b$, where $S_f$ is the concentration of free scaffold and $K_a$ and $K_b$ are equilibrium dissociation constants.  Note that $K_a = k_{-a}/k_a$ and $K_b = k_{-b}/k_b$, where $k_a$, $k_{-a}$, $k_b$ and $k_{-b}$ are rate constants in the reaction scheme of Fig.~\ref{fig:1}(D). Thus,  
\begin{equation}
V^{\prime} =\frac{h k_{\rm cat} B_0A_f}{K_m^{\prime}+A_f} \ ,
\label{eq:scf_kinetics}
\end{equation}
where $K_m^{\prime} = K_a(1+K_b/S_f)$.  If $h$ in Eq.~(\ref{eq:scf_kinetics}) is approximately 1 (i.e., scaffold effects on the reactivity of scaffold-tethered enzyme and substrate are inconsequential, which can be the case \cite{Park:03}), then the effect of a scaffold on the rate of an enzyme-catalyzed reaction lies in the difference between $K_m$ and $K_m^{\prime}$, the apparent Micheaelis-Menten constant for the scaffold-mediated reaction.  Note that $K_m^{\prime}$ is unrelated to $K_m$ and is determined by the affinities of the enzyme and substrate for binding sites on the scaffold and by the concentration of free scaffold.  Thus, a scaffold can profoundly affect the catalytic efficiency of an enzyme even if the stereochemical effects of the scaffold on reactivity (captured in $h$) are inconsequential.

\section{Numerical Results}\label{sec:res}

\begin{figure}
\centering
\includegraphics[scale=0.33]{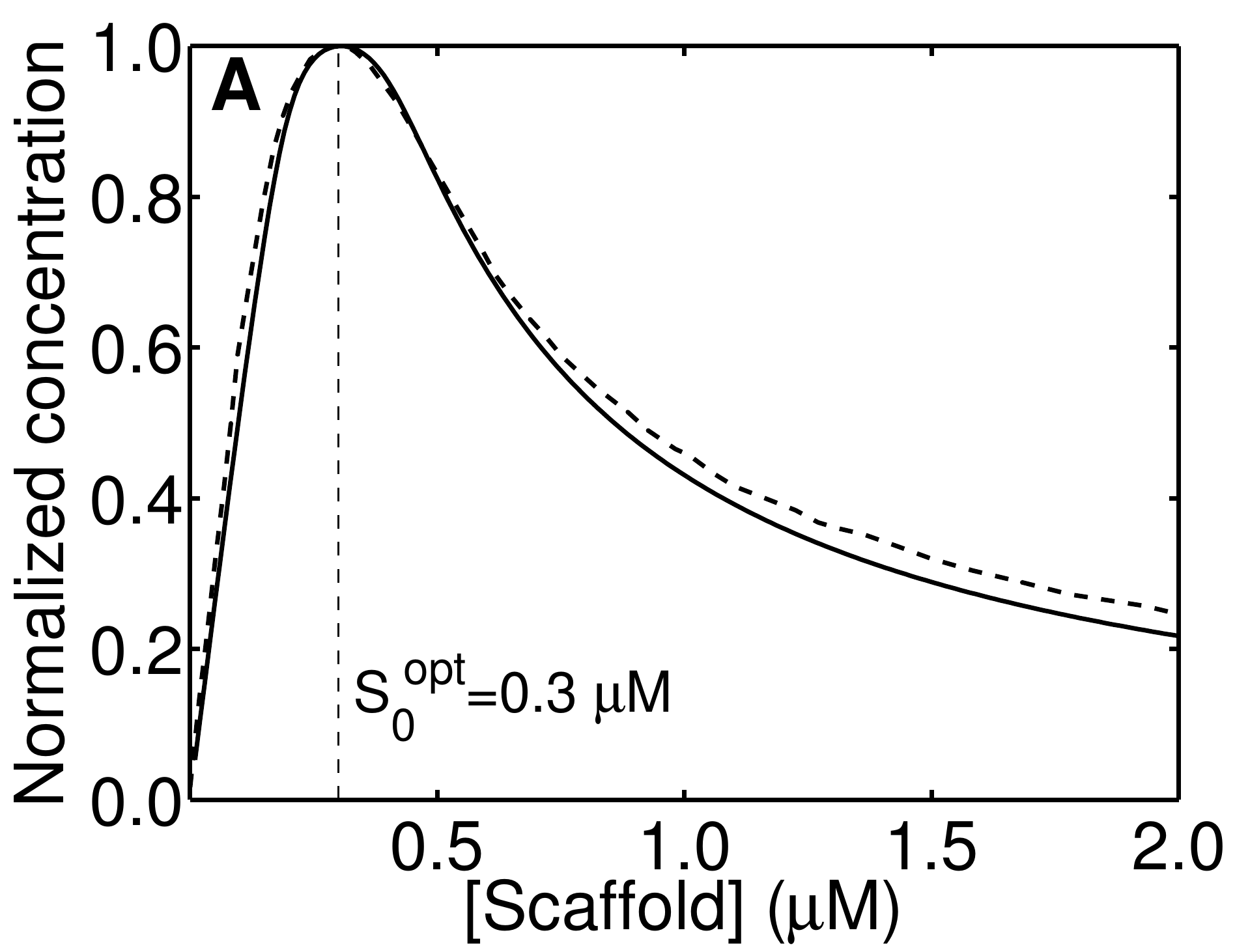} \\
\hspace{-0.06in}\includegraphics[scale=0.33]{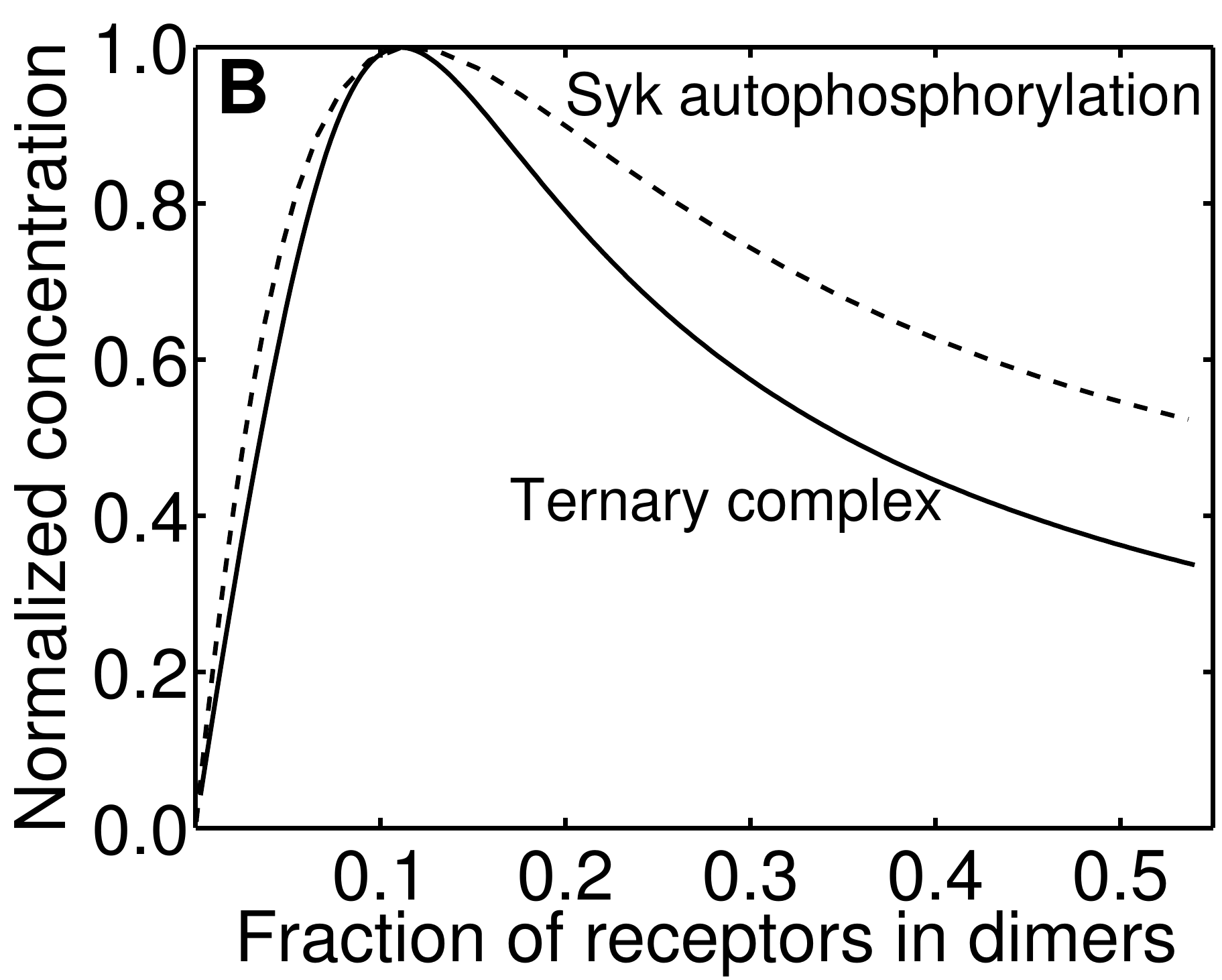}
\caption{Comparison of predictions of the ternary-complex model and more mechanistically detailed models capturing the effects of scaffolds. (A) Comparison between the results of the
ternary-complex model (solid line) and a model for a MAP kinase cascade incorporating a scaffold \cite{lev:00}. The broken
line corresponds to the concentration of  phosphorylated
MAPK in Ref.~\cite{lev:00}. Parameter values correspond to those used
in Ref.~\cite{lev:00}: $A_0=0.2$ $\mu$M, $B_0=0.4$ $\mu$M and $K_a=K_b=5$ nM. (B)
Comparison between the results of the ternary-complex model (solid line) and a model for early events in IgE receptor signaling \cite{goldstein2002,FaederJamesR.:InveeF}. The broken line corresponds to the amount
of Syk autophosphorylation predicted by the model of Ref.~\cite{FaederJamesR.:InveeF}
for the case of $4\times 10^4$ Syk molecules per cell, as in Fig. 3
of Ref.~\cite{hlavacek2003ccs}. Other parameter values are as
in Ref.~\cite{FaederJamesR.:InveeF}. Note that the $K_D$ for Syk binding to
phosphorylated receptor is 2167 molecules per cell.}  
\label{fig:2}
\end{figure}

\subsection{Relevance of simple ternary-complex model}

To investigate scaffold-mediated nucleation of signaling complexes, we
will focus, except as noted, on a bivalent scaffold that recruits a pair of monovalent ligands, i.e., on our ternary-complex model given by Eqs.~\ref{eq:sa}--\ref{eq:b0} and illustrated in Fig.~\ref{fig:1}(B). To demonstrate the relevance of this minimalist model, we will consider two more mechanistically detailed models for cell signaling systems reported in
the literature. The first of these models characterizes the effect of a scaffold on a MAP
kinase cascade~\cite{lev:00}. The second model characterizes early events of
immunoreceptor signaling, in which a ligand-induced receptor dimer nucleates a
signaling complex~\cite{goldstein2002,FaederJamesR.:InveeF}.

In the model of Levchenko et al.~\cite{lev:00}, MAPK is phosphorylated as a
result of a cascade of binding and phosphorylation reactions in which a scaffold
co-localizes MAPK with its kinase, MAPKK. In Fig.~\ref{fig:2}A, we plot the
steady-state level of phosphorylation of MAPK predicted by the model
of Levchenko et al.~\cite{lev:00} as a function of scaffold concentration.  In addition, we plot
the dependence of scaffold-nucleated ternary complex concentration ($C_{ab}$) on scaffold concentration using the parameters that characterize the binding
reactions of MAPK, MAPKK and scaffold in the original model. As can be
seen, if we assume that MAPK phosphorylation is related to the amount of MAPK
and MAPKK co-localized by scaffold, then our simple model closely predicts the
steady-state level of MAPK phosphorylation predicted by the model of Levchenko et
al.~\cite{lev:00} even though the simple model omits many of the reactions
included in the more mechanistically detailed model. This result suggests that
the steady-state output of the MAP kinase cascade is dominated by co-localization of
MAPK and MAPKK by the scaffold. We note that the MAPK-scaffold and MAPKK-scaffold
equilibrium dissociation constants in the model of~\cite{lev:00} are each 5 nM.  As a result, $S_0^{\rm opt}$ is largely determined by the concentrations
of MAPK (0.4 $\mu$M) and MAPKK (0.2 $\mu$M), as can be seen from
Eq.~(\ref{eq:optim}).  

In the mechanistic model for early events in IgE receptor (Fc$\epsilon$RI) signaling developed by Goldstein et al.~\cite{goldstein2002} and Faeder et
al.~\cite{FaederJamesR.:InveeF}, a bivalent ligand induces receptor 
dimerization; receptor phosphorylation; recruitment of Syk, a cytosolic protein
tyrosine kinase, to phosphorylated receptors;  and autophosphorylation of
receptor-associated Syk. In this model, a phosphorylated receptor dimer acts as
a scaffold that co-localizes two copies of Syk, enabling one copy of Syk to
phosphorylate its neighboring copy of Syk.  In Fig.~\ref{fig:2}B, we plot the
steady-state level of Syk autophosphorylation predicted by the detailed model with the parameters of Ref.~\cite{FaederJamesR.:InveeF}
as a function of receptor dimer concentration. We also plot the dependence of $C_{ab}$ on phosphorylated receptor dimer (the effective scaffold) concentration using the
parameters of Ref.~\cite{FaederJamesR.:InveeF} that characterize the binding reactions of phosphorylated receptor
and Syk. As can be seen, if we
assume that Syk autophosphorylation is related to the number of receptor dimers
associated with two copies of Syk, then the ternary-complex model accurately predicts the
amount of ligand-induced receptor dimers that yields maximum Syk
autophosphorylation. This result further demonstrates that the minimalist
ternary-complex model can provide insights into the effects of a scaffold on
cell signaling. 

We caution that the ternary-complex model is not guaranteed to
recapitulate the system properties of more sophisticated, more mechanistic models.
 Our point is only that the model, despite its simplicity, can be
useful for gaining insights into signaling events in which scaffold-mediated
nucleation plays a role, as illustrated by the qualitative similarities of the two sets of curves in Fig.~\ref{fig:2}.  The ternary-complex model can fail to capture the effects of a scaffold in the context of a specific cell signaling system for myriad reasons.  For example, an important caveat of the model is that it is an equilibrium model.  Thus, if interactions of binding partners with a scaffold are slow relative to the time scale of processes affected by scaffold-mediated nucleation, then the ternary-complex model will fail to capture the true effects of the scaffold, which will be dominated by the kinetics of ligand-scaffold interactions.  Also, the scope of the ternary-complex model is rather circumspect.  For example, it does not account for the process of Syk dephosphorylation in the model of Refs.~\cite{goldstein2002,FaederJamesR.:InveeF}, or more generally, downstream influences on events triggered by scaffold-mediated nucleation. This feature of the ternary-complex model (e.g., omission of processes affecting Syk phosphorylation status, including dephosphorylation mediated by phosphatases) explains the quantitative discrepancy between the two curves of Fig.~\ref{fig:2}(B).  

\begin{figure}
\begin{center}
\includegraphics[scale=0.33]{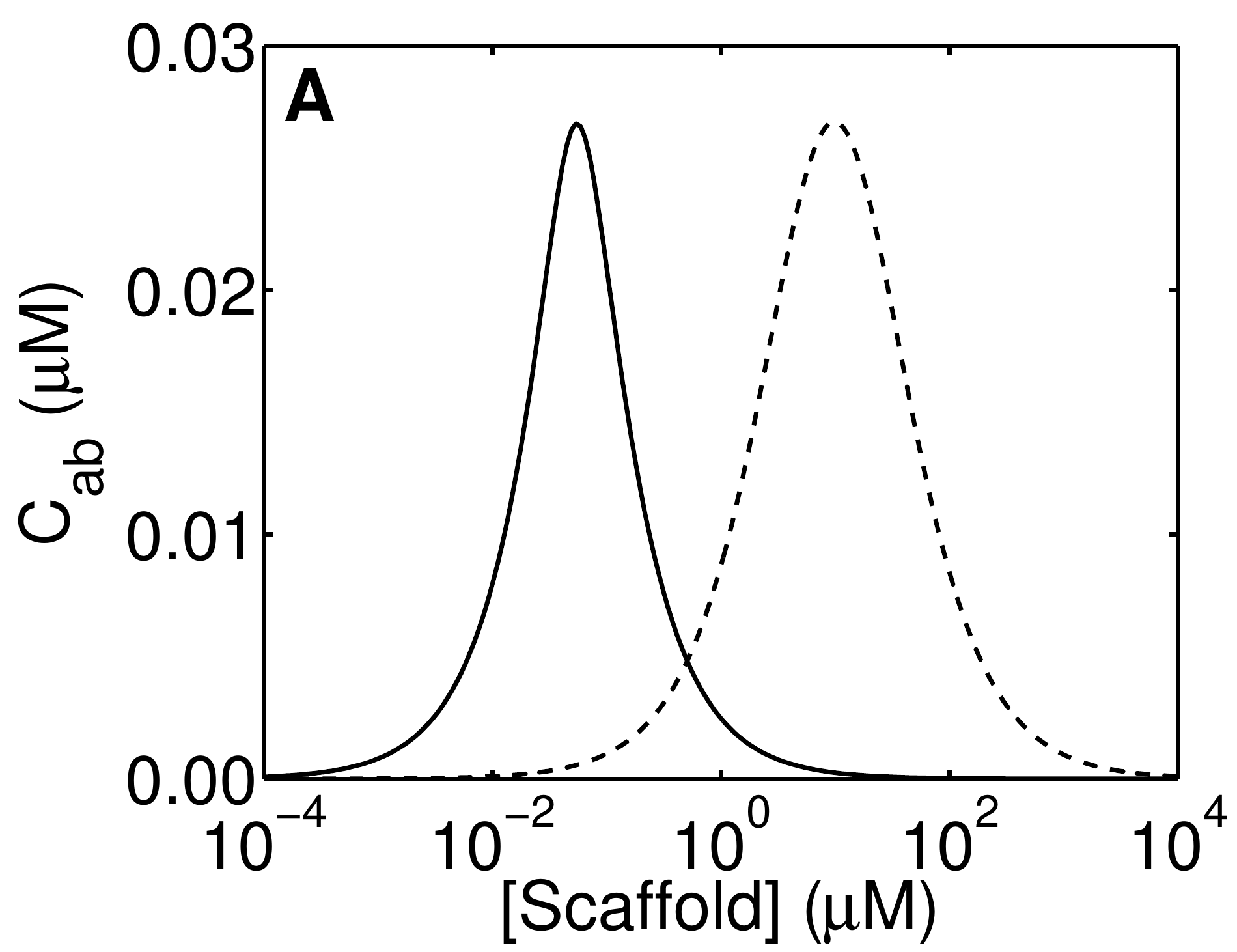}
\includegraphics[scale=0.9]{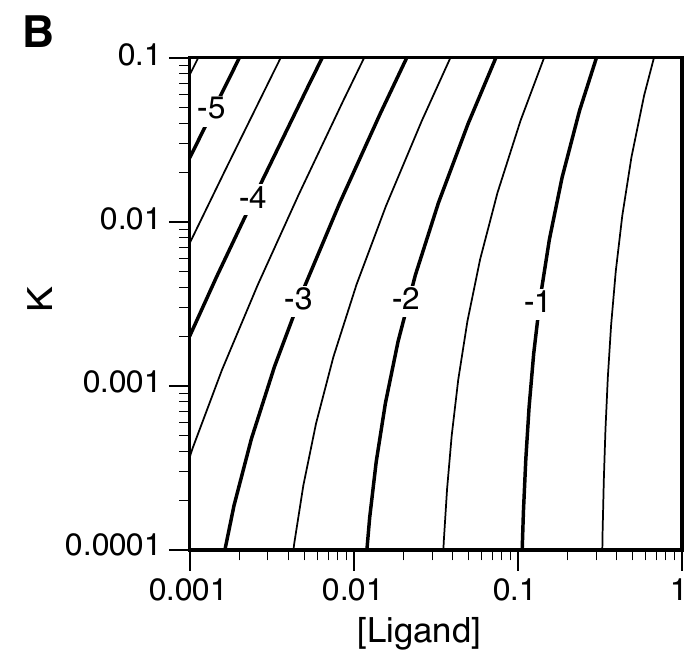}
\end{center}
\caption{A scaffold is potentially multifunctional.  (A) According to the ternary-complex model, a scaffold can recruit different ligand pairs at different scaffold concentrations. Parameters used in
calculations are as follows: $A_0=B_0=0.05$ $\mu$M and $K_a=K_b=5$ nM (solid
line);  $A_0=B_0=1.0$ $\mu$M and $K_a=K_b=8.8$ $\mu$M (broken line). Competition of ligand pairs was not considered explicitly in our calculations because competition is insignificant for this particular example.  (B) A contour plot showing $\log_{10} (C_{ab}^{\rm max}/ 1 \mbox{ } \mu{\rm M})$ as a function of ligand concentration $L_0$ ($\mu$M) and equilibrium dissociation constant $K_D$ ($\mu$M).  We assume that $K_a=K_b=K_D$ and $A_0=B_0=L_0$.  Along any contour line, $C_{ab}^{\rm max}$ is constant while $S_0^{\rm opt}$ varies in accordance with Eq.~(\ref{eq:optim}).} \label{fig:3} 
\end{figure}

\subsection{Concentration-dependent scaffold functionality}

A bivalent scaffold, depending on its concentration, can recruit different pairs
of ligands. In Fig.~\ref{fig:3}A, we show that two ligand
pairs can be recruited to the same scaffold and to the same maximal extent at
different scaffold concentrations. Co-localization of the first ligand pair is
maximum at a scaffold concentration of 0.055 $\mu$M, whereas co-localization of
the second ligand pair is maximum at 9.8 $\mu$M.  In general, such a scenario
requires that the ligand pairs have different total concentrations and/or scaffold
affinities (Fig.~\ref{fig:3}B). The contour plot of Fig.~\ref{fig:3}B shows how $C_{ab}^{\rm max}$, the concentration of ternary complex at $S_0=S_0^{\rm opt}$, depends on ligand concentration and affinity for the special case where $A_0=B_0=L_0$ and $K_a=K_b=K_D$.  Thus, at different concentrations, a scaffold can produce quantitatively identical outputs, as measured by co-localization of its binding partners, but qualitatively distinct outputs, because the binding partners differ.  Any scaffold is potentially multifunctional, as the concentration of a scaffold can be adjusted through regulation of gene expression. In addition, the effective concentration of a scaffold can be adjusted
without changes in gene expression through post-translational modifications,
e.g., phosphorylation-dependent binding sites on a scaffold can be turned on and
off by kinases and phosphatases, effectively changing the amount of scaffold
available for interaction with ligands. 

\begin{figure}
\begin{center}
\includegraphics[scale=0.33]{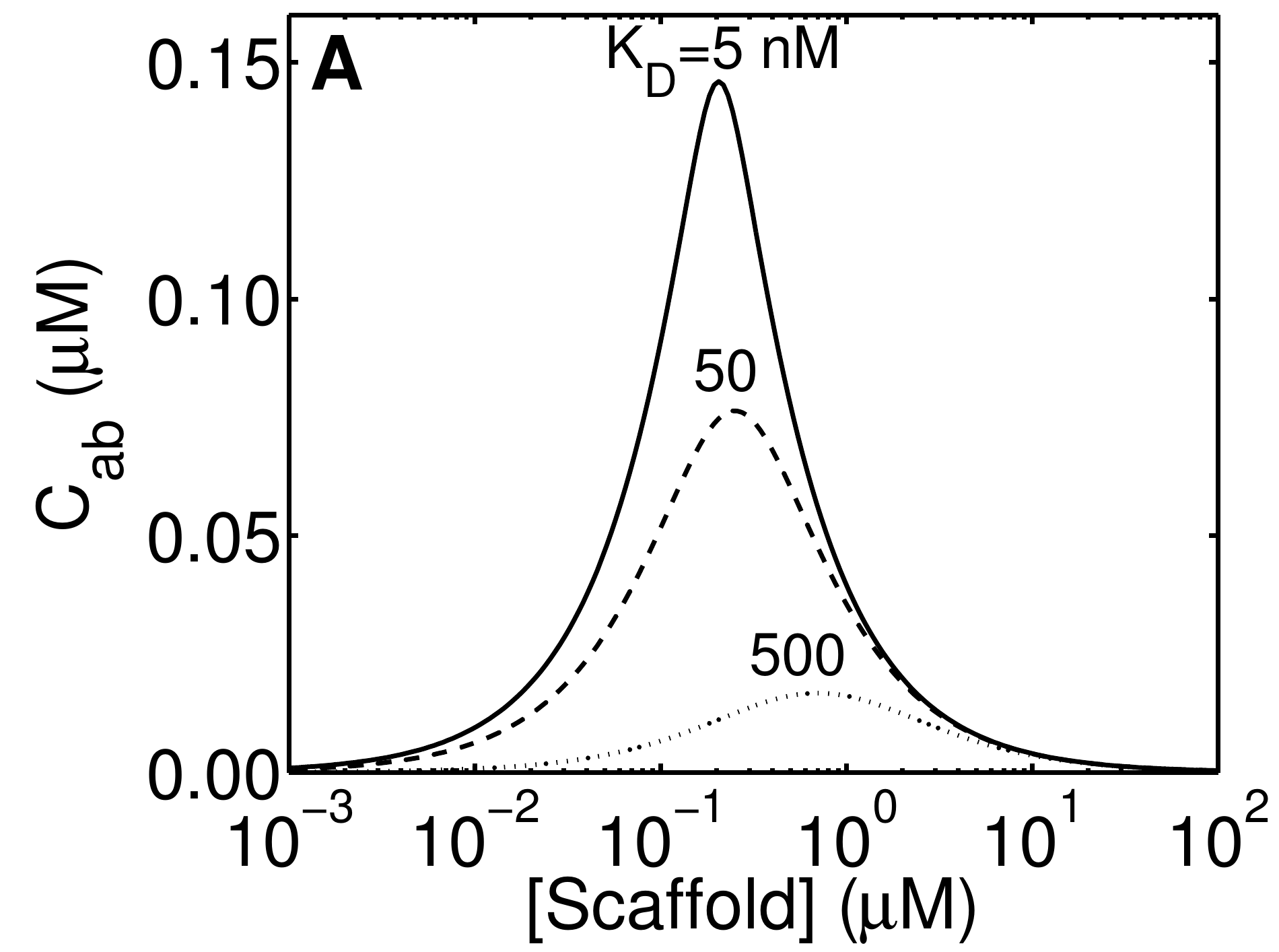}\\
\includegraphics[scale=0.33]{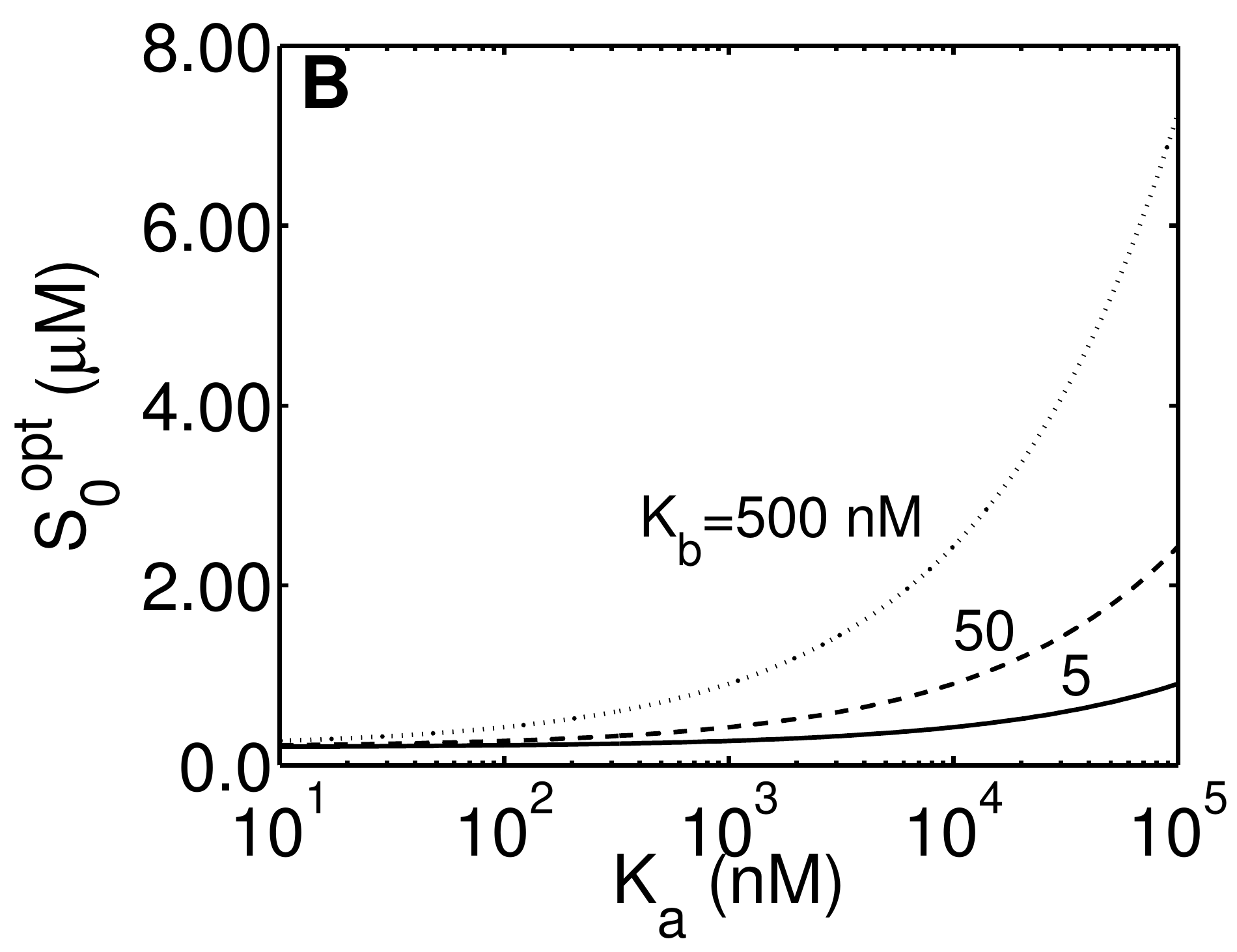}\\
\includegraphics[scale=0.33]{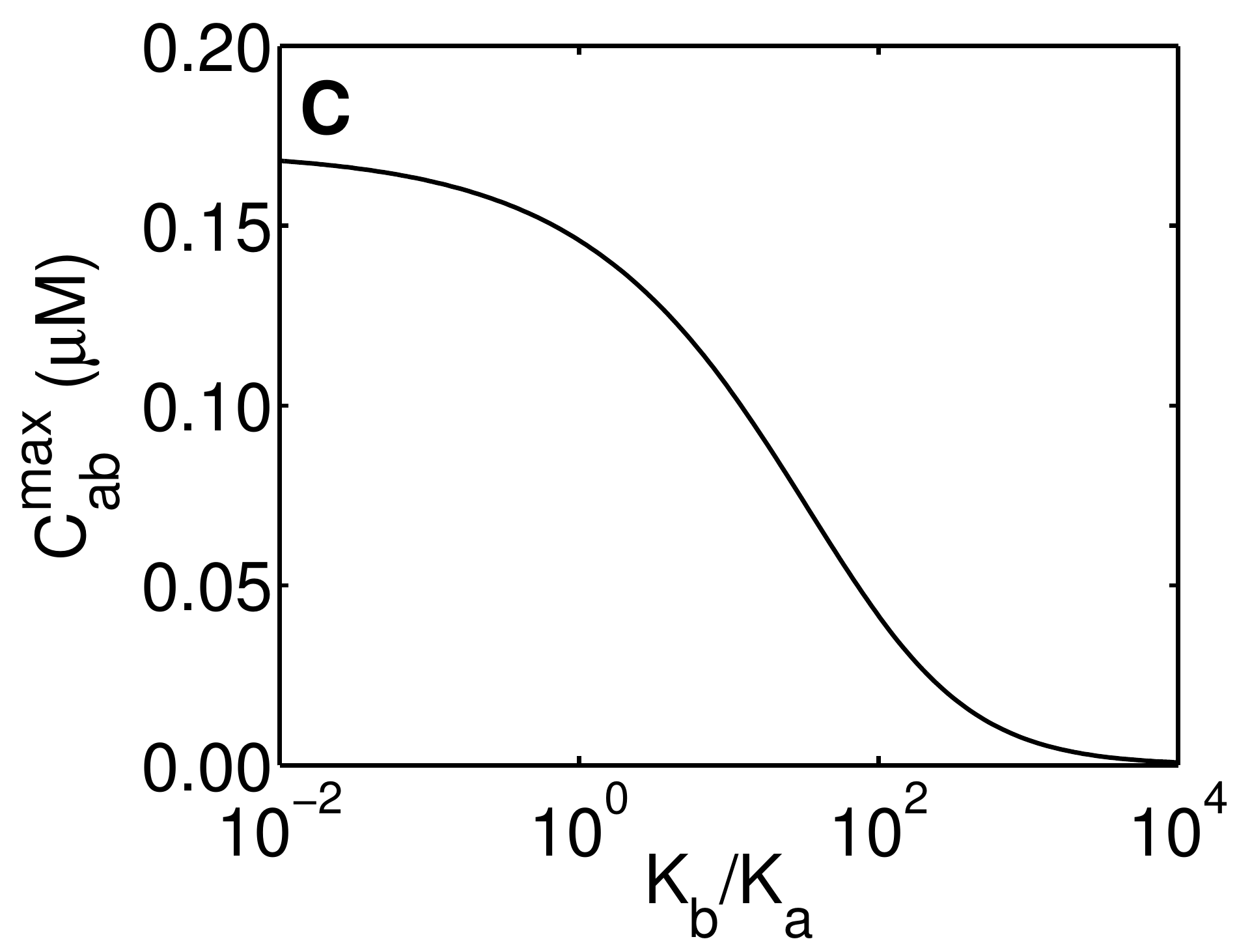}
\end{center}
\caption{Effects of binding constants on optimal scaffold concentration
$S_0^{\rm opt}$ and maximum concentration of ternary complex
$C_{ab}^{\rm max}$.  (A) The value of $C_{ab}$ is plotted as the function of
$S_0$ for three different values of $K_D$.  In these calculations, $K_a$ and $K_b$ are each set equal to $K_D$ and $A_0=B_0=0.2$ $\mu$M. (B) The value of $S_0^{\rm opt}$ is shown as a function of
$K_a$ for three different values of $K_b$. Ligand concentrations are the same as
for panel (A).  (C) The value of $C_{ab}^{\rm max}$ is plotted as a function of the
ratio of the equilibrium dissociation constants ($K_b/K_a$). In this panel, $K_a$
is held fixed at 5 nM and ligand concentrations are the same as for panel
(A).}\label{fig:4} 
\end{figure}

\subsection{Dependence of complex formation on binding constants}

Figure~\ref{fig:4} illustrates how the maximum concentration of ternary complex, $C_{ab}^{\rm max}$, and the scaffold concentration at which $C_{ab}=C_{ab}^{\rm max}$ ($S_0^{\rm opt}$) depend on the binding constants that characterize ligand-scaffold interactions.  In Fig.~\ref{fig:4}A, we can see
that $S_0^{\rm opt}$ increases as the equilibrium dissociation constant $K_D$ ($K_a=K_b=K_D$ for the results shown in this panel) increases, or equivalently, as affinity decreases.  In Fig. \ref{fig:4}B, we can see how $S_0^{\rm opt}$ changes
as  the equilibrium dissociation constant for scaffold interaction with ligand $A$ ($K_a$) is varied  while $K_b$ is held
constant at 5, 50, and 500 nM. The value of $S_0^{\rm opt}$ is least sensitive to changes in $K_a$ and $K_b$ for small values of $K_a$ and $K_b$ (i.e., for high ligand-scaffold affinities).
Figure~\ref{fig:4}C shows how $C_{ab}^{\rm max}$ depends on $K_a$ and $K_b$. In this example, $C_{ab}^{\rm
max}$ is affected by changes of $K_a$ and $K_b$, but large changes of $K_a$ or
$K_b$ result in only modest changes of $C_{ab}^{\rm max}$. For example, a 100-fold
increase in $K_{b}$ from 5 nM ($K_b/K_a=1$) to 0.5 $\mu$M ($K_b/K_a=100$)
results in a only an approximate four-fold decrease of $C_{ab}^{\rm max}$ from 0.15 $\mu$M to 0.04
$\mu$M.  As indicated by Eq.~(\ref{eq:app}), the sensitivities of $C_{ab}^{\rm max}$ and $S_0^{\rm opt}$ to an increase in a binding constant is analogous to the sensitivities of these quantities to an increase in the concentration of a competitive inhibitor or decoy receptor.

\begin{figure}
\begin{center}
\includegraphics[scale=0.33]{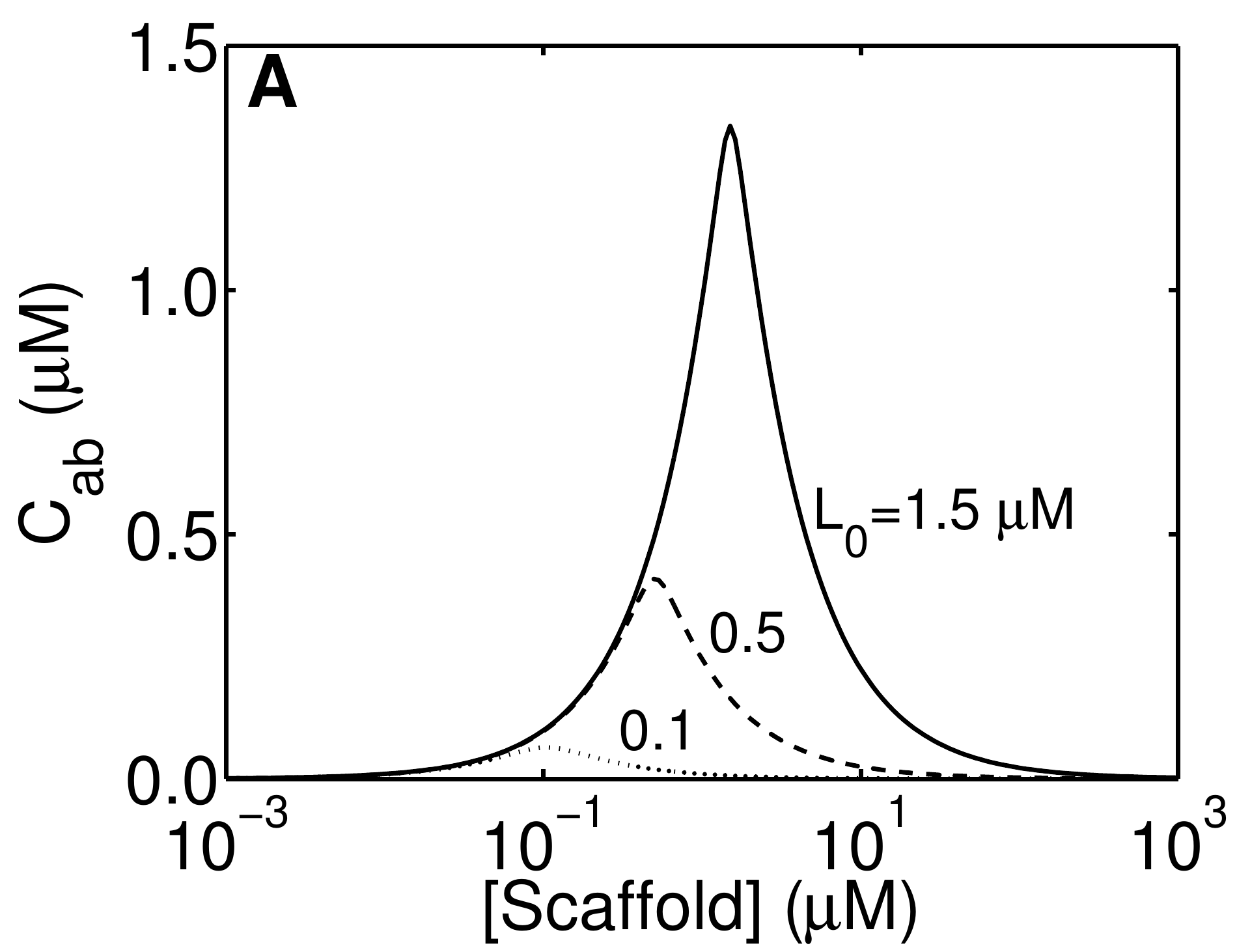}
\includegraphics[scale=0.33]{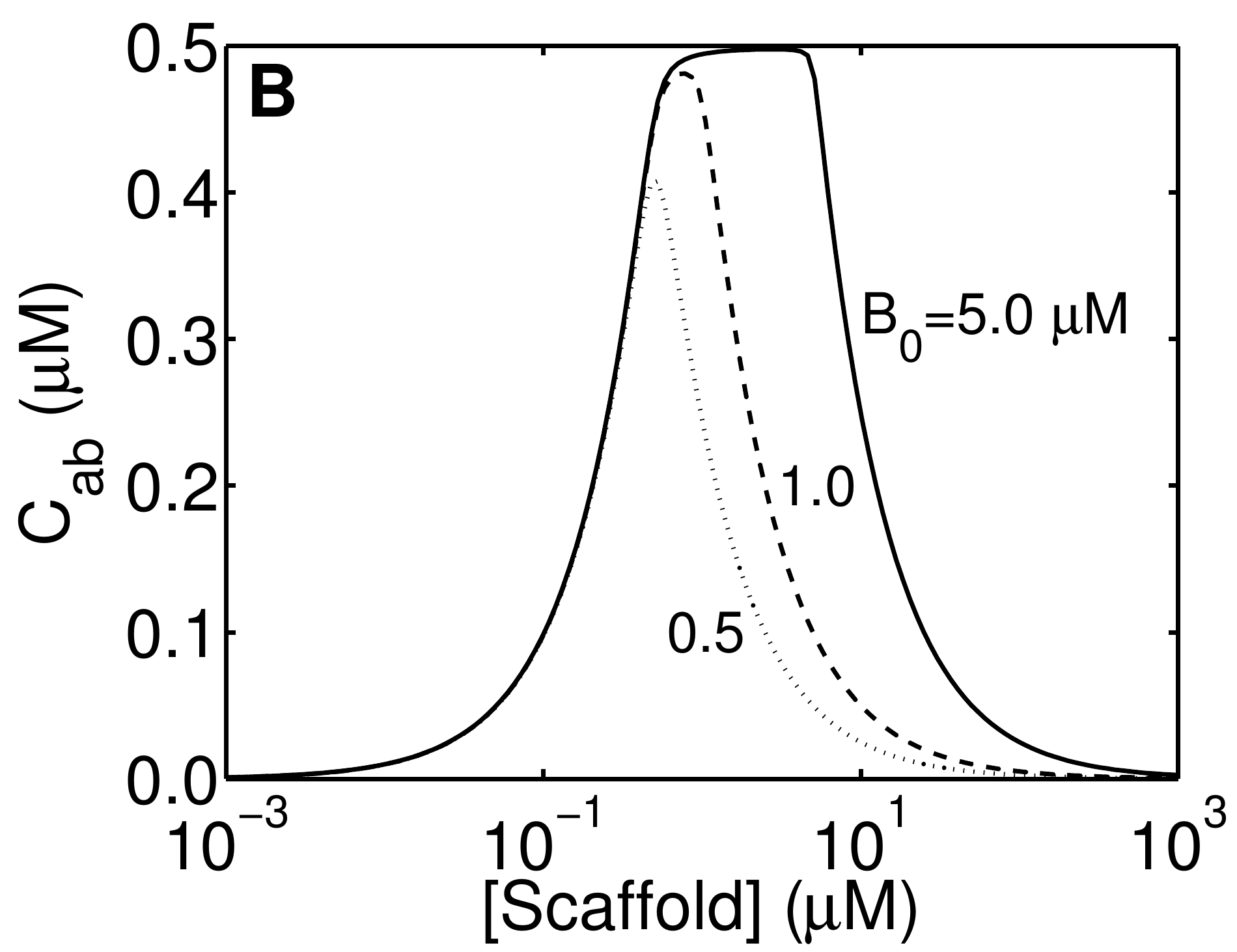}
\end{center}
\caption{\label{fig:5} Dependence of scaffold-nucleated complex formation on
ligand concentrations according to the ternary-complex model. (A) The amount of
ternary complex $C_{ab}$ is plotted as a function of total scaffold
concentration $S_0$ for three different values of $L_0$.  In these calculations, $A_0$ and $B_0$ are each set equal to $L_0$ and $K_a=K_b=5$ nM. 
(B) $C_{ab}$ vs. $S_0$ for three different total concentrations of ligand $B$ ($B_0$) with $A_0$ held fixed at
0.5 $\mu$M. The equilibrium dissociation constants are the same as for panel (A).} 
\end{figure}

\subsection{Dependence of complex formation on ligand concentrations}

Figure~\ref{fig:5} illustrates how the maximum concentration of ternary complex, $C_{ab}^{\rm max}$, and the scaffold concentration at which $C_{ab}=C_{ab}^{\rm max}$ ($S_0^{\rm opt}$) depend on ligand concentrations.  The results shown in Fig.~\ref{fig:5} complement those of Eq.~(\ref{eq:optim}) and Fig.~\ref{fig:3}. In Fig.~\ref{fig:5}A,  we can see that $S_0^{\rm
opt}$ increases as both ligand concentrations ($A_0=B_0=L_0$) increase
simultaneously. Increasing ligand concentrations also increases $C_{ab}^{\rm max}$.
Thus, inhibition of ternary complex formation by excess scaffold can be reversed
by increasing ligand concentrations. For example, 10 $\mu$M scaffold inhibits
formation of the ternary complex at $L_0=0.5$ $\mu$M. In contrast, when
$L_0=1.5$ $\mu$M, $C_{ab}^{\rm max}$ is increased by nearly three fold and there is a significant amount of ternary complex at a 10 $\mu$M concentration of scaffold
(Fig.~\ref{fig:5}A).  In Fig.~\ref{fig:5}B, we show how $C_{ab}$ depends on the total scaffold concentration
$S_0$ for three different values of $B_0$ (0.5, 1 and 5 $\mu$M). For all three
curves, the value of $A_0$ is the same (0.5 $\mu$M). As can be seen, when the
two ligand concentrations are different, the value of $C_{ab}$ can be nearly
maximal over a broad range of scaffold concentrations. For example, at $B_0=5.0$
$\mu$M and $A_0=0.5$ $\mu$M, $C_{ab}$ is nearly maximal for scaffold
concentrations between 1 and 5 $\mu$M. Over this range, ligand $B$ is in excess
and ligand $A$ is limiting (i.e, nearly all $A$ is bound to scaffold and the
amount of ternary complex is approximately equal to the concentration of ligand
$A$). At higher scaffold concentrations, free $B$ is consumed and $C_{ab}$
becomes less than maximal because of the excess scaffold effect~\cite{lay:97,lev:00}.  

\begin{figure}
\begin{center}
\includegraphics[scale=1.0]{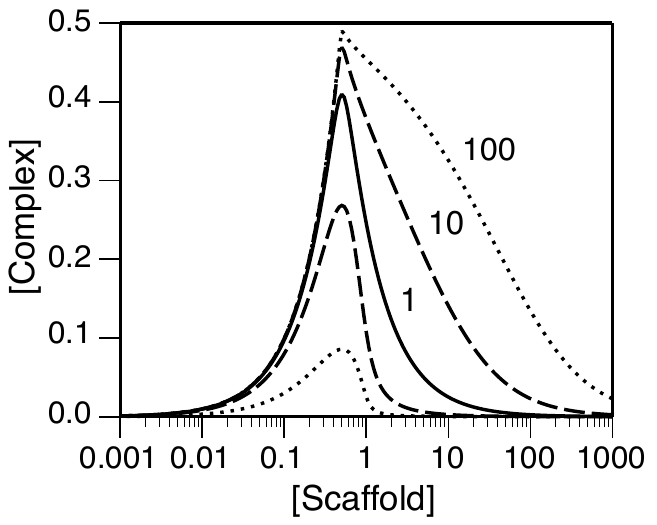}
\end{center}
\caption{The effect of cooperativity on scaffold-mediated nucleation of ternary complex: $C_{ab}$ ($\mu$M) is plotted as a function of $S_0$ ($\mu$M) for five different values of the cooperativity factor $\phi$.  The solid line corresponds to independent binding of ligands $A$ and $B$ to the scaffold ($\phi=1$).  The broken line and dotted line \textit{below} the solid line correspond to negativity cooperativity: $\phi=0.1$ and $\phi=0.01$, respectively.  Similarly, the broken line and dotted line \textit{above} the solid line correspond to positive cooperativity: $\phi=10$ and $\phi=100$, respectively.  Calculations are based on the following ligand concentrations and binding constants:
$A_0=B_0=L_0=0.5$ $\mu$M and $K_a=K_b=K_D=5$ nM.  In accordance with Eq.~(\ref{eq:optim}), $S_0^{\rm opt}=L_0+K_D=0.505$ $\mu$M for all curves.} 
\label{fig:6}
\end{figure}

\subsection{Cooperative binding}

Cooperative binding of ligands $A$ and $B$ to a scaffold is illustrated in
Fig.~\ref{fig:6}.  In accordance with Eq.~(\ref{eq:optim}), $S_0^{\rm opt}$ is
independent of the value of $\phi$. However, for $\phi>1$, $C_{ab}$ takes on a
nearly maximal value over a broader range of scaffold concentrations. For
$\phi<1$, the maximal value of $C_{ab}$ is attenuated.   

\begin{figure}
\begin{center}
\includegraphics[scale=0.33]{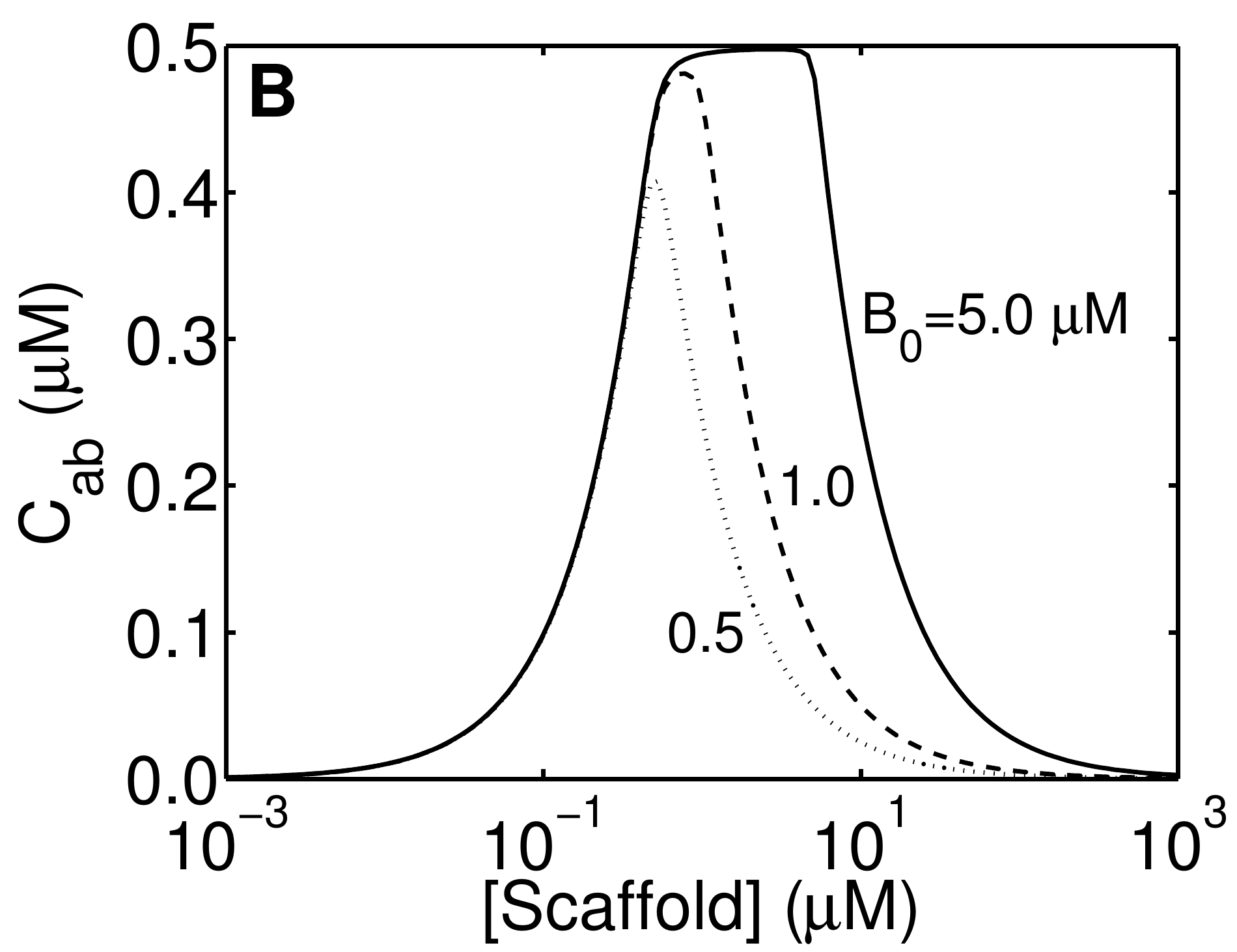}
\end{center}
\caption{The effects of scaffold valence $n$ on nucleation of a complex containing $n$ ligands ($L_1,\ldots,L_n$). The concentration of complex containing all $n$ ligands, $C_n$, is plotted as a function of the total  scaffold concentration, $S_0$, for three different valences: $n=2$, 3 and 4. Calculations are based on Eqs.~(\ref{eq:bc}) and (\ref{eq:complex}) and the following parameter values: $L_{i_0}=1$ $\mu$M  and $K_i=0.05$ $\mu$M for all $i$.} \label{fig:7}
\end{figure}

\subsection{Higher-order complexes}

So far, we have considered ternary complexes nucleated by a bivalent scaffold. Here, we consider scaffold-nucleated complexes containing two or more ligands per scaffold. In other words, we consider the reaction scheme of Eq.~\ref{eq:1}, which is illustrated in Fig.~\ref{fig:1}A. In Fig.~\ref{fig:7}, we show how $C_2$, $C_3$ and $C_4$ depend on the total scaffold concentration $S_0$, where $C_n$ is the concentration of scaffold bound to $n$ ligands. As can be seen, for a larger value of $n$, the maximum value of $C_n$ is reduced (in  a controlled comparison). In addition, the scaffold concentration at which $C_n$ is maximum is reduced. These results are consistent with the intuition that it is more difficult to form a scaffold-nucleated complex that contains more ligands. The results of Fig.~\ref{fig:7} are essentially the same as those obtained by Heinrich et al.~\cite{hein:02} through the analysis of a more mechanistically detailed model, which provides further support for the relevance of the minimalist models considered in this study.  The asymmetry seen in the curves of Fig.~\ref{fig:7} is explained by the asymptotic limits of $|g_n|$, which is given by Eq.~(\ref{eq:dcds}), as $S_0$ approaches 0 and $\infty$.

\begin{figure}[ht]
\center\includegraphics[scale=1.0]{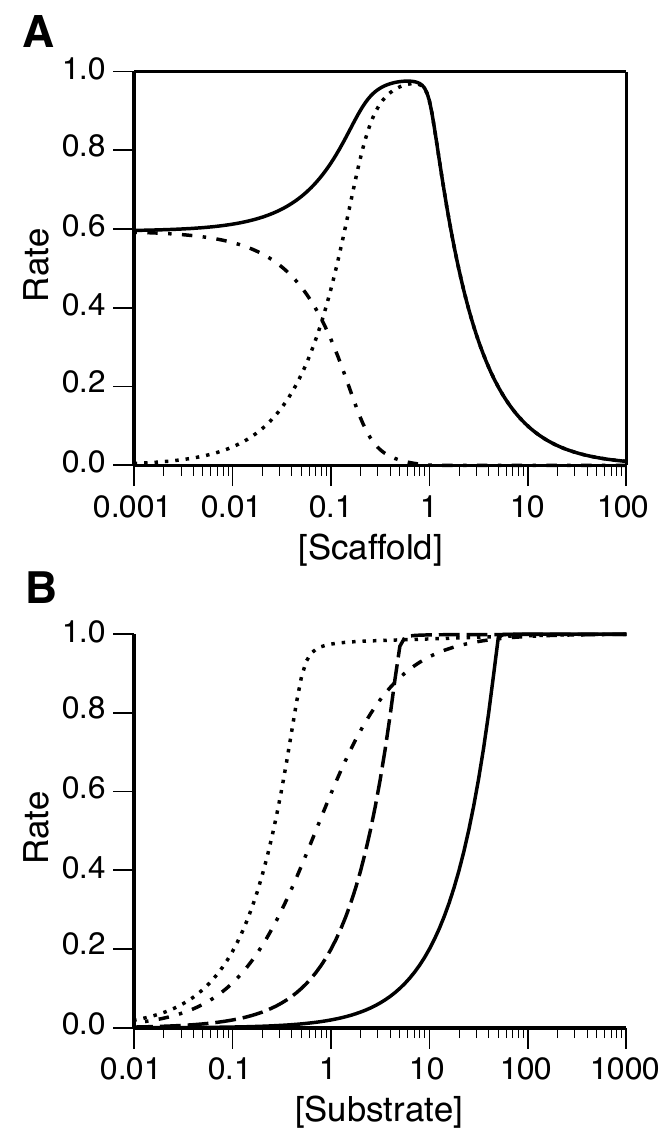}
  \caption[The effect of a scaffold on an enzymatic reaction]{\label{fig:rate}
The effect of a scaffold on an enzymatic reaction under the assumption of rapid equilibrium.  The initial normalized rate of reaction is plotted as a function of (A) total scaffold concentration $S_0$ ($\mu$M) and (B) total substrate concentration $A_0$ ($\mu$M).  In panel (A), $A_0=1$ $\mu$M.  The dash-dot curve corresponds to $V/(k_{\rm cat}B_0)$ (Eq.~(\ref{eq:MM})), the dotted curve corresponds to $V^{\prime}/(hk_{\rm cat}B_0)$ (Eq.~(\ref{eq:scf_kinetics})), and the solid curve corresponds to $V/(k_{\rm cat}B_0)+V^{\prime}/(hk_{\rm cat}B_0)$. In panel (B), the dash-dot curve corresponds to the case of a classical Michaelis-Menten reaction (no scaffold).  The other curves correspond to different scaffold concentrations: 0.5 $\mu$M (dotted line), 5 $\mu$M (broken line) and 50
$\mu$M (solid line). Calculations are based on Eqs.~(\ref{eq:sa})--(\ref{eq:b0}) with $\phi=1$ and with Eqs.~(\ref{eq:a0}) and (\ref{eq:b0}) each modified to include a term for the concentration of $(AB)$, which is taken to be $A_fB_f/K_m$.  The following parameter values were used in all calculations: $K_a=K_b=0.005$ $\mu$M, $K_m=0.6$ $\mu$M, and $B_0=0.2$ $\mu$M.} 
\end{figure}

\subsection{Effects of a scaffold on an enzymatic reaction}

In accordance with Eq.~(\ref{eq:scf_kinetics}), an enzyme-scaffold complex can be regarded as an effective enzyme, one that has a different maximal rate of reaction and one that achieves the half maximal rate at a different substrate concentration, which can be lower or higher than the corresponding concentration in the absence of scaffold. Figure~\ref{fig:rate} complements these insights, showing how a scaffold that co-localizes a substrate and enzyme can affect the initial rate of enzyme-catalyzed conversion of the substrate to product.  As can be seen in Fig.~\ref{fig:rate}A, a scaffold can significantly accelerate the rate of reaction relative to the situation where the reaction proceeds in the absence of the scaffold.  However, scaffold-mediated acceleration occurs only within a certain range of scaffold concentrations.  A scaffold concentration outside this range, in the excess scaffold regime, can have the effect of inhibiting the enzyme-catalyzed reaction.  In Fig.~\ref{fig:rate}B, the rate of reaction is shown as a function of substrate concentration for different concentrations of scaffold and for the case of no scaffold. As can be seen, the dose-response curve can be shifted by the involvement of a scaffold, and in addition, the slope of the curve can be altered, i.e., the sensitivity of reaction rate to a change in substrate concentration can be altered.  For the case of Fig.~\ref{fig:rate}B, the effect of the scaffold is to increase the sensitivity of the initial reaction rate to a change in substrate concentration.

\section{Discussion}\label{sec:discuss}

Elucidating the design principles of cellular regulatory systems is recognized as an important goal of systems biology \cite{savageau2001,wall2004,alon2006}, which is facilitated by the tools of synthetic biology \cite{mukherji2009,bashor2010}.  Understanding the design principles of scaffold-mediated nucleation of a multicomponent complex, a motif in cell signaling \cite{good2011}, is important for a number of reasons, including understanding how scaffold function is conserved across species, interpreting the functional consequences of protein copy number variation in diseases, and engineering scaffolds to have desired properties \cite{Li10}.

Here, we have used minimalist models for which analytical results can be obtained to reveal principles that govern the recruitment of ligands to a scaffold. We have shown that the simplest model considered here, the ternary-complex model, captures essential features predicted by more sophisticated and mechanistically detailed models, which can only be analyzed via simulation (Fig.~\ref{fig:2}). For the the ternary-complex model, we derived a number of analytical expressions, which can serve as basic design equations, in that these equations can potentially be used to guide the manipulation of intrinsic scaffold properties or scaffold milieu to achieve desired effects on the system-level properties of scaffold-dependent processes.  Perhaps the most significant analytical result is Eq.~(\ref{eq:optim}).  Special cases of this equation have been reported in studies of multivalent ligand-receptor binding.  For example, Mack et al.~\cite{mack2008exact} obtained a similar equation in a study of bivalent ligand-receptor binding (cf. Eq. (16) in Ref.~\cite{mack2008exact} and Eq.~(\ref{eq:optim}) here in this report).  In studies specifically focused on scaffolds, earlier work has largely relied on simulation.  The parameter-dependent nature of simulation results can lead to incomplete understanding.  For example, on the basis of simulation results, Levchenko et al. \cite{lev:00,levchenko2004rmg} reported that scaffold function is insensitive to binding constants, which is definitely the case in certain parameter regimes.  However, as can be seen from Eq.~(\ref{eq:optim}) and as illustrated in Figs.~\ref{fig:3} and \ref{fig:4}, binding constants can have a significant impact on scaffold function.  

For the ternary-complex model, in our presentation of Eqs.~(\ref{eq:sa})--(\ref{eq:optim}), we have shown that the maximum concentration of the ternary complex, $C_{ab}^{\rm max}$, and the scaffold concentration that yields this maximum concentration, $S_0^{\rm opt}$, depend on ligand-scaffold binding constants and ligand concentrations (Eq.~(\ref{eq:optim})).  In addtion, $C_{ab}^{\rm max}$, but not $S_0^{\rm opt}$, depends on the cooperativity of ligand $A$ and ligand $B$ binding to the scaffold.  Levchenko et al. \cite{lev:00} reported that cooperativity shifts the optimal scaffold concentration (see Fig.~4B in Ref.~\cite{lev:00}).  This finding is inconsistent with the equilibrium binding properties of a scaffold (Eq.~\ref{eq:optim}).  The shift of dose-response curves seen in the study of Levchenko et al. \cite{lev:00} is perhaps attributable to kinetic binding properties of a scaffold in the specific context of a MAP kinase cascade. The above mentioned analytical results are complemented by the simulation results of Figs.~\ref{fig:3}--\ref{fig:6}.  In our presentation of Eqs.~(\ref{eq:1})--(\ref{eq:concave}), we have shown that scaffold-mediated nucleation depends asymmetrically on total scaffold concentration, which is illustrated in Fig.~\ref{fig:7}.  This finding is consistent with the simulation results of Heinrich et al. \cite{hein:02} (see Fig.~8 in Ref.~\cite{hein:02}).  We also considered competitive ligands and decoy receptors, the presence of which can be mimicked by an increase in the equilibrium binding constants for ligand-scaffold interactions (Eq.~(\ref{eq:app})).  Finally, we found that a scaffold can reprogram the catalytic efficiency of an enzyme (Eq.~(\ref{eq:scf_kinetics}) and Fig.~\ref{fig:rate}).

The results of Fig.~\ref{fig:3}A suggest that scaffolds could have multiple
context-dependent functions. A scaffold, depending on its concentration, can
nucleate distinct sets of ligands. Thus, in a given cellular context, only a
subset of a scaffold's known binding partners may be physiologically relevant.
In addition, in cases where the (effective) scaffold concentration depends on
the strength of a signal, a scaffold may enable a cell to respond differentially
to signal strength. An example where effective scaffold concentration depends on
signal strength via post-translational modifications and other fast events (and
not gene expression) is provided by the model of Goldstein et al.
\cite{goldstein2002} and Faeder et al.~\cite{FaederJamesR.:InveeF} for signaling
by the high-affinity receptor for IgE antibody (Fc$\epsilon$RI). In this model,
phosphorylated receptors in ligand-induced receptor dimers serve to co-localize
two copies of the protein tyrosine kinase Syk, which is necessary for Syk
autophosphorylation and downstream Syk-dependent signaling events. Furthermore,
the number of receptor dimers depends on the concentration of (multivalent)
ligand that induces receptor aggregation. Thus, given the results of
Fig.~\ref{fig:3}A, receptor dimers in this system, which have a scaffold-like
function, could nucleate different signaling complexes at different ligand doses
\cite{hlavacek2003ccs,nag2010}, with these complexes potentially eliciting
different cellular responses. We are not aware of a signaling system where
scaffold concentration controls the formation of complexes that lead  to
distinct cellular responses, but given the number of binding partners of a
typical signaling protein, we expect that  scaffold multifunctionality, such as
that illustrated in Fig.~\ref{fig:3}A, could be found. 

Results such as those shown in Fig.~\ref{fig:4}C can be used to engineer a scaffold with non-native
ligand-scaffold interactions to obtain a desired scaffold-dependent cellular
response to a signal, assuming ternary complex formation correlates with the
cellular response of interest. Park et al.~\cite{Park:03} engineered the yeast
scaffold protein Ste5 to interact with one of its binding partners, Ste7 or
Ste11, via a non-native protein-protein interaction. In these experiments, the
Ste7 or Ste11 binding site on Ste5 was disrupted by a mutation of Ste5
(V763A/S861P or I504T) and Ste5 was fused to the PDZ domain of syntrophin. In
addition, Ste7 or Ste11 was fused to the PDZ domain of neuronal nitric oxide
synthase (nNOS). The nNOS and syntrophin PDZ domains heterodimerize with a $K_D$
of 0.6 $\mu$M~\cite{har:01}. In contrast, the native interaction between Ste5
and Ste7 or Ste11 is characterized by a $K_D$ of about 0.1
$\mu$M~\cite{maeder2007srf,slaughter2007mdp}. Using measured concentrations of Ste5 (35 nM), Ste11 (39 nM) and Ste7 (68 nM) reported by Maeder et al.~\cite{maeder2007srf}, we can calculate the effect of increasing $K_D$ from 0.1 to 0.6 $\mu$M. We find that this change is expected to cause a
15-fold decrease in the amount of the ternary complex composed of Ste5, Ste7 and
Ste11. This prediction is consistent with the decrease in phosphorylation of
Fus3  observed by Park et al.~\cite{Park:03} for the engineered Ste5-Ste7 or
Ste5-Ste11 interaction.  

We have focused on equilibrium properties of scaffolds that nucleate ternary and higher-order protein complexes.  We also considered the effect of a scaffold on the rate of an enzyme-catalyzed reaction under a rapid equilibrium assumption.  The results that we have found should be useful for understanding cellular responses that depend on the concentration and composition of a scaffold-nucleated complex, assuming that these responses are insensitive to the kinetics of complex formation. As Locasale et al.~\cite{locasale2007spc,locasale2008rsd} and Thalhauser and Komarova~\cite{thalhauser2010} have shown, the kinetics of scaffold-mediated complex formation
can have physiological relevance. Thus, an important caveat of the results presented here is the assumption of equilibrium or pseudo equilibrium.  Another caveat is the assumption of monovalent ligands.  If the ligands that interact with a multivalent scaffold are themselves multivalent, then complex dose-response behavior can arise, such that complex formation is maximal at multiple scaffold concentrations \cite{hlavacek2003ccs,nag2010}.  

An interesting future application of the results presented here might be interpretation of comparative proteomics data.  It has been observed that protein copy number tends to be conserved across species \cite{schrimpf2009,weiss2010,laurent2010}.  Because the function of a scaffold is affected by the relative abundances of the scaffold and its ligands, conservation of relative abundances of the proteins in a ``scaffold motif'' would suggest conservation of function of the motif, whereas the opposite would suggest the acquisition of a new function or compensatory modifications of scaffold properties and/or milieu, which could perhaps be more easily understood or predicted in light of this report.

\section*{Acknowledgements}

This work was supported by NIH grants GM076570 and RR18754, Department of Energy
contract DE-AC52-06NA25396, and NSFC grant 30870477. J.Y. thanks Dr. Zhen Wang
for helpful discussions and the Center for Nonlinear Studies for support that
made travel to Los Alamos possible.

\end{document}